\documentclass[twocolumn]{scrartcl}
\usepackage{epsfig}
\usepackage{cite,latexsym,amsmath,caption}
\begin{document}
\title{Photodetachment study of He$^{-}$ quartet resonances below 
the He(n=3) thresholds}
\author{A E Klinkm\"uller$^{\$}$, G Haeffler$^{\$}$, D Hanstorp$^{\$}$,
    I Yu Kiyan$^{\$}$ \thanks{Permanent address: Russian Academy of
    Sciences, General Physics Institute, Vavilova St.~38, 117\,942
    Moscow, Russia }~ ,\\ U Berzinsh$^{\$}$ \thanks{Permanent address:
    Institute of Atomic Physics and Spectroscopy, University of Latvia,
    LV 1586 Riga, Latvia }~ and D J Pegg$^{\#}$ \\ $^{\$}$Department of
    Physics, Chalmers University of Technology\\ and G\"oteborg
    University, SE-412\,96 G\"oteborg, Sweden \\ $^{\#}$Department of
    Physics, University of Tennessee, Knoxville,\\ Tennessee 37\,996,
    USA}
\date{\today}
\maketitle
\begin{abstract}
The photodetachment cross section of He$^{-}$ has been measured in
the photon energy range $2.9\ldots 3.3$~eV in order to investigate
doubly excited states.  Measurements were made channel specific by
selectively detecting the residual He atoms left in a particular
excited state following detachment. Three Feshbach resonances were
found in the He(1s2p\,$^{3}$P$^{\text{o}}$)+e$^{-}$($\epsilon$p)
partial cross section: a $^{4}$S resonance below the
He(1s3s\,$^{3}$S) threshold and two $^{4}$P resonances below the
He(1s3p\,$^{3}$P$^{\text{o}}$) threshold. The measured energies of
these doubly excited states are 2.959\,260(6)~eV, 3.072(7)~eV and
3.264\,87(4)~eV. The corresponding widths are found to be
0.20(2)~meV, 50(5)~meV and 0.61(5)~meV\@. The measured energies
agree well with recent theoretical predictions for the
1s3s4s\,$^{4}$S, 1s3p$^{2}$\,$^{4}$P and 1s3p4p\,$^{4}$P states,
respectively, but the widths deviate noticeably from calculations
for 1s3p$^{2}$\,$^{4}$P and 1s3p4p\,$^{4}$P states.
\end{abstract}
\begin{flushleft}
\end{flushleft}
\section{Introduction}
Doubly excited states of two-electron atomic systems provide us with an
opportunity to investigate the interplay of electron-electron and
electron-core interactions. Negative ions are of particular interest
because here the normally dominant Coulomb interactions between the
electrons and the core are suppressed, thus enhancing the role of the
interelectronic interaction. Pioneering photodetachment experiments on
the pure two-electron negative ion H$^{-}$ were first performed by
Bryant and co-workers \cite{Bry-77}. Several experiments have been
performed on Li$^{-}$ \cite{Ber-95-1,Lju-96}, which can be considered an
effective two-electron ion since the two valence electron move in a
spherically symmetric potential created by the inert Li$^{+}$ core. The
three-electron He$^{-}$ ion is also an effective two-electron system at
low photon energies since the 1s core electron remains inert up to
energies of 40~eV and only slightly perturbs the motion of the two
active outer electrons.

The He$^{-}$ ion is the prototype of an unusual class of negative ions
that are not stable, but rather metastable, against decay via
autodetachment. The lowest energy state of this ion is the
1s2s2p\,$^{4}$P$^{\text{o}}$ state, which is bound by 77.516(6)~meV
relative to the 1s2s\,$^{3}$S state of He \cite{Kri-97-2}. This state
can autodetach into the ground 1s$^{2}$\,$^{1}$S state of the He atom
via the relatively weak magnetic interactions. The $J$=$5/2$ level, for
example, has a measured lifetime of 350(15)~$\mu$s \cite{And-93}, making
the He$^{-}$ ion sufficiently long lived to survive a transit through a
typical ion beam apparatus.

There have been several previous experimental investigations of the
photodetachment of He$^{-}$ \cite{Com-80,Hod-81,Pet-85,Peg-90,Wal-94-4,
Kli-97} at photon energies in the visible. Calculations of the
He$^{-}$ photodetachment cross section and resonance parameters have
also been made \cite{Haz-81,Sah-90,Dou-90,Dav-90,The-94,The-95,Xi-96,
Byl-97,Kim-97}. In this paper we present the results of a recent
photodetachment study of double excitation in the He$^{-}$ ion in the
energy range $2.9\ldots 3.3$~eV. We have observed one resonance below
the He(1s3s\,$^{3}$S) threshold and two resonances below the
He(1s3p\,$^{3}$P$^{\text{o}}$) threshold. We have identified the
resonances with the 1s3s4s\,$^{4}$S, 1s3p$^{2}$\,$^{4}$P and
1s3p4p\,$^{4}$P states by comparing their measured energies with the
results of a recent MCHF calculation of Xi and Froese Fischer
\cite{Xi-96}. The 1s3s4s\,$^{4}$S state has been studied previously in a
different partial channel of photodetachment \cite{Kli-97}.

Doubly excited states of He$^{-}$ have also been studied in electron
impact experiments on helium targets
\cite{Bru-77,Buc-83,Buc-87,Buc-94}. Their presence is manifested as
resonance structure in scattering cross sections. Many resonance states
of doublet symmetry have been observed as transient intermediate states
in the electron scattering process. Data on quartet states, however, are
sparse. Such states appear as resonances in the cross section of
photodetachment from the ground quartet state of the He$^{-}$ ion. The
energy resolution in photodetachment experiments is typically much
higher than in electron scattering experiments. Thus, the energies and
widths of quartet states can, in principle, be measured more accurately
than the corresponding doublet states. Selection rules on
phototexcitation of He$^{-}$ from the 1s2s2p\,$^{4}$P$^{\text{o}}$
ground state limit, however, the study of doubly excited states to those
with $^{4}$S, $^{4}$P and $^{4}$D symmetry.

\section{Experiment}
\label{expe}
\subsection{Procedure}
The present measurements were made using a collinear laser-ion beam
apparatus.  The experimental method has been previously used in studies
of partial photodetachment cross sections for He$^{-}$ and Li$^{-}$
\cite{Hae-96-1,Lju-96,Kli-97}.  It consists of several
sequential steps. A doubly excited state of a negative ion is first
produced by photoexcitation from its ground state. The excited ion then
autodetaches leaving the residual atom, in general, in an excited
state. By selectively detecting those atoms left in a particular state
one can isolate the corresponding decay channel and investigate the
partial cross section. The excited residual atoms are selected by use of
the method of resonance ionization spectroscopy, i.e.~they are further
photoexcited to a Rydberg state which is subsequently field ionized in a
static electric field. The resulting positive ion signal is proportional
to the partial cross section for the selected decay channel.

\begin{figure}
\begin{center}
{\epsfig{file=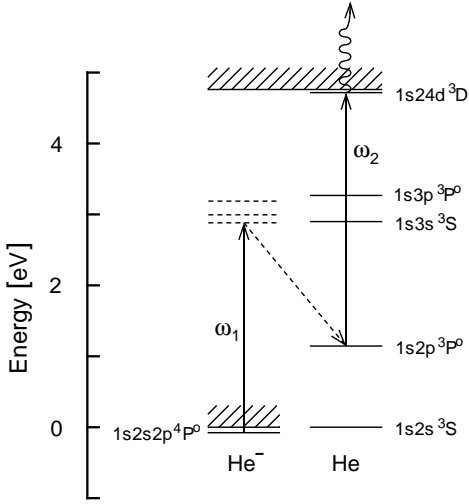, width=0.86\columnwidth}}
\end{center}
\protect\caption{\label{kli-fig01}\sloppy Excitation scheme: selected
states of He/He$^{-}$. The horizontal dashed lines show the positions of
the three measured doubly excited states. The diagonal dashed line
indicates autodetachment of one of these states via the
He(1s2p\,$^{3}$P$^{\text{o}}$)+e$^{-}$($\epsilon$p) channel and the wavy
line represents electric-field ionization. }
\end{figure}
The investigation of doubly excited states of He$^{-}$ in the present
experiment is based on measurements of resonance structure in the
partial photodetachment cross section for the
He(1s2p\,$^{3}$P$^{\text{o}}$)+e$^{-}$($\epsilon$p) channel. The
excitation-detection scheme used is shown in figure \ref{kli-fig01}. In
the first step, metastable He$^{-}$ ions are photodetached in a laser
field of frequency $\omega_{1}$:
\begin{multline*}
\text{He}^{-}(\text{1s2s2p}\,^{4}\text{P}^{\text{o}}) + \hbar\omega_{1}
\mapsto\\ \text{He}(\text{1s2p}\,^{3}\text{P}^{\text{o}}) +
\text{e}^-(\epsilon\text{p}) \quad .
\end{multline*}
If, in addition, the above process proceeds via an intermediate doubly
excited state of He$^{-}$, resonance structure will appear in the
photodetachment cross section. This structure which, in general, has an
asymmetric shape, is the result of an interference between the resonant
and non-resonant photodetachment processes.

In the second step, He atoms, produced in the 1s2p\,$^{3}$P$^{\text{o}}$
state, as a result of photodetachment, are resonantly excited into the
24d state by a laser field of frequency $\omega_{2}$, sufficiently
strong to saturate the transition. These Rydberg atoms are subsequently
ionized in a static electric field of about
200~$\frac{\text{kV}}{\text{m}}$:
\begin{equation}\label{eq2}\begin{split}
\text{He}(\text{1s2p}\,^{3}\text{P}^{\text{o}}) + \hbar\omega_{2}
&\mapsto \text{He}(\text{1s24d}\,^{3}\text{D})\quad ,
\\
\text{He}(\text{1s24d}\,^{3}\text{D}) &\leadsto
\text{He}^{+}(\text{1s}\,^{2}\text{S}) + \text{e}^{-}\quad ,
\end{split}
\end{equation}
where $\leadsto$ represents field ionization.

The yield of the He$^{+}$ ions produced in this state-selective
detection scheme was recorded as a function of the frequency
$\omega_{1}$, while the frequency $\omega_{2}$ was held constant on
the transition to the Rydberg state. The signal was proportional to
the population of helium atoms left in the 1s2p\,$^{3}$P$^{\text{o}}$
state after the photodetachment step. Since the intensity of laser
$\omega_{2}$ and the ion beam current were constant during a scan, the
He$^{+}$ signal, normalized to the intensity of laser $\omega_{1}$,
was proportional to the
He(1s2p\,$^{3}$P$^{\text{o}}$)+e$^{-}$($\epsilon$p) partial
photodetachment cross section. We experimentally determined that this
signal changed linearly with respect to the power of laser
$\omega_{1}$, thus excluding the possibility of higher order
processes.

The detection scheme, based on the selective detection of residual
He(1s2p\,$^{3}$P$^{\text{o}}$) atoms, was effective in eliminating a
potential background source arising from He atoms produced in the ion
beam via autodetachment of the metastable ground state of
He$^{-}$. These atoms, created in the 1s$^{2}$\,$^{1}$S ground state,
were not photoionized by laser $\omega_{2}$ .

\subsection{Experimental arrangement}
The He$^{-}$ beam was produced from a mass-selected $^{4}$He$^{+}$ ion
beam via charge exchange in a Cs vapor cell. The beam energy was 3.1~keV
and the ion current was typically 1 nA.

\begin{figure}
\begin{center}
\epsfig{file=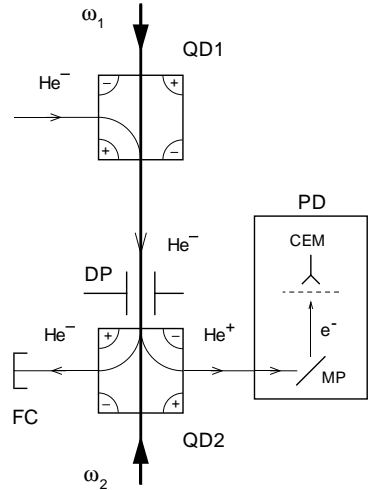, width=0.66\columnwidth}
\end{center}
\protect\caption{\label{kli-fig03}\sloppy The interaction and detection
regions: \textsf{QD1,QD2}, electrostatic quadrupole deflectors;
\textsf{CEM}, channel electron multiplier; \textsf{DP}, deflection
plate; \textsf{PD}, positive ion detector; \textsf{FC}, Faraday cup;
\textsf{MP}, metal plate.}
\end{figure}

In the interaction region, shown schematically in figure
\ref{kli-fig03}, the laser and ion beams were coaxially superimposed
over a distance of 50~cm between the two electric quadrupole deflectors
(\textsf{QD1, QD2}).  The beam paths were defined by apertures of 3~mm
diameter at both ends of the interaction region. The apparatus has
previously been described in more detail \cite{Han-95,Kli-97-4}.

One source of background of He$^{+}$ ions arises from double detachment
collisions of He$^{-}$ with the residual gas.  This contribution was
reduced significantly by installing a pair of deflection plates
(\textsf{DP}) just before the second quadrupole deflector
(\textsf{QD2}). The transverse electric field between the deflection
plates was insufficient to field ionize the Rydberg atoms, but strong
enough to sweep collisionally-created He$^{+}$ ions out of the beam. To
monitor the He$^{-}$ beam current, the deflection plates were
periodically grounded.

The highly excited Rydberg atoms of He were field ionized by the static
electric field of the second quadrupole deflector (\textsf{QD2}). The
resulting He$^{+}$ ions were, in turn, deflected by this quadrupole into
the positive ion detector (\textsf{PD}), where they impinged on a metal
plate (\textsf{MP}) and produced secondary electrons. These electrons
were detected with a channel electron multiplier (\textsf{CEM}).

The two laser frequencies $\omega_{1}$ and $\omega_{2}$ used in the
experiment were produced by two dye lasers pumped by a common XeCl
excimer laser that delivered pulses of about 15~ns duration. The laser
light of frequency $\omega_{2}$ was generated by PTP dye with a pulse
energy of typically 1~mJ\@. The tunable laser light of frequency
$\omega_{1}$ was generated by Exalite 416, Exalite 411 and BBQ dyes with
pulse energies of typically 1~mJ\@. The signal was normalized to the
power of laser $\omega_{1}$.

For the measurement of the narrow resonances associated with the
1s3s4s\,$^{4}$S and 1s3p4p\,$^{4}$P states the frequency $\omega_{1}$
was calibrated by combining Fabry-Perot fringes with reference lines
generated in a hollow cathode lamp. The Fabry-Perot fringes served as
frequency markers whereas the reference transitions in argon and uranium
provided an absolute calibration of the energy scale. In
the measurement of the broad resonance associated with the
1s3p$^{2}$\,$^{4}$P state the energy scale was calibrated using our
measured position of the 1s3s4s\,$^{4}$S state situated only
887~cm$^{-1}$ away. In this case a relative scale was established by use
of the reading of the laser which has an accuracy of better than
0.03~cm$^{-1}$ \cite{Kli-97-4}.

\section{Results and discussion}
\label{con}
\begin{figure}
\begin{center}
\epsfig{file=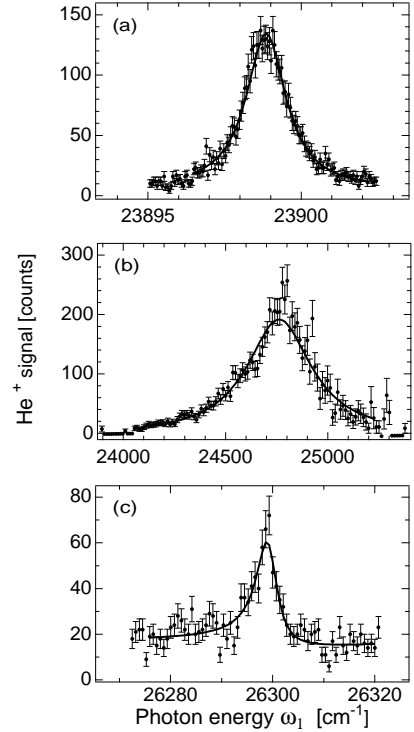,width=0.75\columnwidth}
\end{center}
\protect\caption{\label{kli-fig05}\sloppy Yield of He$^{+}$ ions
vs. photon energy $\omega_{1}$. The solid line is a fit to the
experimental data (dots) using equation \eqref{eq6}. The three curves
show the (a) 1s3s4s\,$^{4}$S; (b) 1s3p\,$^{2}$\,$^{4}$P and (c)
1s3p4p\,$^{4}$P resonances. The figures show the measured data without
corrections for the Doppler shift (see text). Note that the energy scale
in (a), (b) and (c) is different. }
\end{figure}

Typical spectral scans of the three resonances in the
He(1s2p\,$^{3}$P$^{\text{o}}$)+e$^{-}$($\epsilon$p) partial cross
section are shown in figure \ref{kli-fig05}.

In the case of the 1s3s4s\,$^{4}$S resonance spectrum shown in figure
\ref{kli-fig05}(a) there is a small background due mainly to the
population of the 1s2p\,$^{3}$P$^{\text{o}}$ state of He by (a)
collisional detachment of He$^{-}$ ions and (b) non-resonant
photodetachment of He$^{-}$ by laser $\omega_{2}$:
\begin{multline*}
\text{He}^{-}(\text{1s2s2p}\,^{4}\text{P}^{\text{o}}) + \hbar\omega_{2}
\mapsto\\
\text{He}(\text{1s2p}\,^{3}\text{P}^{\text{o}}) +
\text{e}^{-}(\epsilon\text{p})\quad .
\end{multline*}
The He atoms formed in the 1s2p\,$^{3}$P$^{\text{o}}$ state by either
(a) or (b) will be resonantly ionized according to the processes
described in equation \eqref{eq2}. The collisional detachment
contribution (a) was reduced by maintaining a pressure of $5\times
10^{-7}$~Pa ($5\times 10^{-9}$~mbar) in the interaction chamber. The
contribution (b) was reduced by strongly attenuating the output of laser
$\omega_{2}$. The output remained sufficiently strong, however, to
saturate the Rydberg transition.

A third source of background (c) can, in principle, make a
contribution. The non-resonant photodetachment of He$^{-}$ via the
He(1s3s\,$^{3}$S)+e$^{-}$($\epsilon$s,d) channels will populate the
1s3s\,$^{3}$S state of He. This state will subsequently radiatively
decay to the 1s2p\,$^{3}$P$^{\text{o}}$ state with a lifetime of 36~ns
\cite{Wie-66}. This process was found to be negligible when induced by
laser $\omega_{2}$ due to the previously mentioned attenuation of the
output of this laser. It can, however, make a contribution when induced
by laser $\omega_{1}$. An exception is the 1s3s4s\,$^{4}$S spectrum
where the photon energy $\omega_{1}$ is insufficient to populate the
1s3s\,$^{3}$S state of the He atom.  The relatively short lifetime of
the 1s3s\,$^{3}$S state makes (c) an efficient process for populating
the 1s2p\,$^{3}$P$^{\text{o}}$ state.  Unfortunately, this contribution
is proportional to the intensity of laser $\omega_{1}$, and attenuation
of this intensity cannot be used to improve the signal-to background
ratio. It, however, remains constant across the scan region of the
spectra shown in figure \ref{kli-fig05}(b) and figure
\ref{kli-fig05}(c).  The signal-to-background ratio is lower for the
1s3p4p\,$^{4}$P spectrum because the peak photodetachment cross section
for this resonance is smaller than for the 1s3p$^{2}$\,$^{4}$P resonance
\cite{Xi-96-2}.

The cross section $\sigma (E)$ in the vicinity of a resonance can be
parametrized as \cite{Sho-68}:
\begin{equation}\label{eq6}
\begin{split}
\sigma (E) &= a+\frac{b\epsilon+c}{1+\epsilon^{2}}
\\
\epsilon &= \frac{E-E_{0}}{(\Gamma /2)}\quad ,
\end{split}
\end{equation}
where $E_{0}$ is the resonance energy, $\Gamma$ is the resonance width,
$E$ is the photon energy, $a$ is the background cross section, and $b,c$
are the Shore parameters. This function is least-square fitted to the
weighted signal and shown as a solid line in figure
\ref{kli-fig05}. More than 20 measured spectra were analyzed for each
resonance. The background was assumed to be constant over the scan
region for all the resonance spectra.

In the case of the narrow 1s3s4s\,$^{4}$S and 1s3p4p\,$^{4}$P
resonances data were recorded with both co- and counter-propagating
laser $\omega_{1}$ and ion beams. The geometric mean of the measured
blue- and red-shifted resonance energies $E^{\text{b,r}}_{0}$ yields
the Doppler-free resonance energy $E_{0}$=$\sqrt{E_{0}^{\text{b}}
E_{0}^{\text{r}}}$. (The energy of the $^{4}$S resonance was
found to be
\begin{equation*}
E_0^{\text{4s}}=23\,868.031(42)\;\text{cm}^{-1}
\end{equation*}
including our previous measurements \cite{Kli-97}, and the
1s3p4p\,$^4$P resonance energy was determined to be
\begin{equation*}
E_0^{\text{4p}}=26\,332.97(40)\;\text{cm}^{-1}\quad .
\end{equation*})
\footnote{The text in brackets is \emph{not} part of the
printed version of this publication.}
For the relatively broad 1s3p$^{2}$\,$^{4}$P resonance, it was
sufficient to record data with counter-propagating laser $\omega_{1}$
and ion beams only. The previously measured 1s3s4s\,$^{4}$S resonance
was, in this case, used to calibrate the laser wavelength scale and to
account for the Doppler shift.

All measured resonance energies and widths are also given in table
\ref{ta06}, together with recent theoretical predictions. In addition,
the experimental values of Klinkm\"uller \textit{et al} \cite{Kli-97}
measured in another partial channel of photodetachment are given. The
quoted uncertainties include both the energy calibration uncertainty and
the statistical scatter of the fitted resonance parameters.

The table shows that there is a good agreement between the measured
and calculated resonance energies.  However there is a noticeable
difference between the measured and calculated widths.  All four
calculations are in essential agreement and all predict the
1s3p$^{2}$\,$^{4}$P state to be narrower than is measured.  In the
case of the 1s3p4p\,$^{4}$P state, the two calculations that predict
the width disagree with each other and both values are larger than the
measured one.

Le Dourneuf and Watanabe \cite{Dou-90} have calculated both doublet and
quartet doubly excited He$^{-}$ resonances below the He(n=3) thresholds
using an adiabatic hyperspherical formalism for the two active
electrons. They found that the He$^{+}$(1s) core only slightly perturbed
the motion of the outer pair of electrons. In fact, the wavefunctions of
the pair displayed correlation patterns very similar to those found in
H$^{-}$. It was suggested that this correspondence allows a
rovibrational classification of the doubly excited states of He$^{-}$.
Many resonance states were labeled in this manner, including two that
could be identified with the 1s3s4s\,$^{4}$S and the 1s3p$^{2}$\,$^{4}$P
states observed in the present experiment. The measured width of the
$^{4}$S state is very small while that of the 1s3p$^{2}$\,$^{4}$P is
large. This observation confirms the radial correlation label "$-$" for
the $^{4}$S state and "$+$" for the $^{4}$P state used by Le Dourneuf
and Watanabe.

\section{Conclusion}
In the present experiment the photodetachment cross section of He$^{-}$
has been studied in the photon energy range $2.9\ldots 3.3$~eV. Three
Feshbach resonances associated with the 1s3s4s\,$^{4}$S,
1s3p$^{2}$\,$^{4}$P and 1s3p4p\,$^{4}$P doubly excited states have been
observed. The measured energy positions of these states agree well with
recent theoretical predictions, but the widths deviate noticeably from
calculations.

We hope to continue this work at higher levels of excitation. It will be
interesting to observe the evolution of states such as
1snpn$^{\prime}$p\,$^{4}$P with increasing values of n, n$^{\prime}$. Of
special interest are the intrashell resonance states 1snp$^{2}$\,$^{4}$P
since it is believed that the ladder of such states with increasing n
leads to double detachment. The 1s2p$^{2}$\,$^{4}$P state appears as a
shape resonance \cite{Wal-94-4} but for n$>$2 the state is bound with
respect to the excited atom and a series of Feshbach resonances appear.
These states are the lowest lying in a manifold of $^{4}$P states below
the excited state thresholds. Bylicki \cite{Byl-97} and Themelis
\textit{et al} \cite{The-94,The-95} have calculated the energies and
widths of these intrashell resonance states up to n=7. Both groups
predict a narrowing of the states with increasing n. Themelis and
Nicolaides show that the configurational composition of these $^{4}$P
states changes significantly with n. For example, in the case of the
present work, the resonance at 3.072~eV photon energy has been labeled
by the dominant configuration, the 1s3p$^{2}$ configuration. This is a
reasonably valid label since, in this case, 1s3p$^{2}$ configuration
amounts to about 79\% of the total composition. As a result of increased
angular correlation, however, the contribution from the 1snd$^{2}$ and
other configurations grow with n. For n$>$4 the 1snp$^{2}$ configuration
no longer dominates and the label becomes invalid. In this manner we can
investigate the gradual breakdown of the independent electron model due
to the increased role of correlation in the determining the motion of
the active pair of electrons.

\section*{Acknowledgements}
J. Xi and C. Froese Fischer are acknowledged for providing us with
unpublished data. Financial support for this research has been provided
by the Swedish National Science Council (NFR). Personal support has been
received for I.K from the Wenner-Gren Center Foundation and for U.B from
the Swedish Institute. D.J.P acknowledges support from the Royal Swedish
Academy of Sciences through itÕs Nobel Institute of Physics and the
U.S. Department of Energy, Office of Basic Energy Sciences, Division of
Chemical Sciences.
\clearpage
{\onecolumn
\begin{table}\small
\caption{\label{ta06} Experimental and theoretical resonance energies
  $E_{0}$ (in eV) and widths $\Gamma$ (in meV). The resonance energies
  were converted from cm$^{-1}$ to eV using the recommended factor
  8\,065.541\,0~$\frac{\text{cm}^{-1}}{\text{eV}}$ \cite{Coh-88}.}
\begin{center}
\begin{tabular}{lr@{.}lr@{.}lr@{.}llr@{.}lr@{.}l}
\hline\hline
& \multicolumn{4}{c}{1s3s4s\,$^{4}$S} &
\multicolumn{3}{c}{1s3p$^{2}$\,$^{4}$P}  
& \multicolumn{4}{c}{1s3p4p\,$^{4}$P}\\
\cline{2-5}\cline{6-8}\cline{9-12} & \multicolumn{2}{c}{$E_{0}$ } &
\multicolumn{2}{c}{$\Gamma$}
& \multicolumn{2}{c}{$E_{0}$ } & \multicolumn{1}{c}{$\Gamma$ }
&\multicolumn{2}{c}{$E_{0}$ } & \multicolumn{2}{c}{$\Gamma$ }\\ \hline
\emph{Experiment}: & \multicolumn{11}{c}{}\\
This work  & 2&959\,260(6) & 0&20(2) & 3&072(7) & 50(5) &
3&264\,87(4) & 0&61(5)\\
Klinkm\"uller \textit{et al}~(1997) \cite{Kli-97} & 2&959\,255(7) &
0&19(3) &
\multicolumn{2}{c}{---} & \multicolumn{1}{c}{---} &
\multicolumn{2}{c}{---} & \multicolumn{2}{c}{---}\\
\emph{Theory}: &  \multicolumn{11}{c}{}\\
Bylicki (1997) \cite{Byl-97} & \multicolumn{2}{c}{---} &
\multicolumn{2}{c}{---} & 3&074\,24 & 37 & 3&264\,78 & 2&45\\ 
Xi \textit{et al}~(1996)  \cite{Xi-96} & \multicolumn{11}{c}{}\\ 
length form & 2&959\,07 & 0&19 & 3&074\,70 & 37.37 & 3&265\,54 &
1&30\\
velocity from & 2&959\,08 & 0&18 & 3&074\,71 & 37.37 & 3&265\,47 &
1&31\\ 
Themelis \textit{et al.}~(1995) \cite{The-95} & \multicolumn{2}{c}{---}
&
\multicolumn{2}{c}{---}  & 3&096\,6 & 34.6 &
\multicolumn{2}{c}{---} & \multicolumn{2}{c}{---}\\
Davis \textit{et al.}~(1990) \cite{Dav-90} &  \multicolumn{2}{c}{---} &
\multicolumn{2}{c}{---}  & 3&086\,8 & 33 &
\multicolumn{2}{c}{---} & \multicolumn{2}{c}{---}\\
\hline\hline
\end{tabular}
\end{center}
\end{table}}
\bibliography{physjabb,qualli}
\end{document}